\documentclass[final, 5p, times, twocolumn]{elsarticle}

\input{preamble.tex}

\journal{arXiv}

\begin{document}

\begin{frontmatter}



\title{Stock Market Telepathy: Graph Neural Networks Predicting the Secret Conversations between MINT and G7 Countries}


\author[label1]{Nurbanu Bursa}

\ead{nurbanubursa@hacettepe.edu.tr}

\affiliation[label1]{organization={Hacettepe University},
            addressline={Department of Statistics}, 
            city={Beytepe},
            postcode={06800}, 
            state={Ankara},
            country={T\"{u}rkiye}}

\begin{abstract}
Emerging economies, particularly the MINT countries (Mexico, Indonesia, Nigeria, and T\"{u}rkiye), are gaining influence in global stock markets, although they remain susceptible to the economic conditions of developed countries like the G7 (Canada, France, Germany, Italy, Japan, the United Kingdom, and the United States). This interconnectedness and sensitivity of financial markets make understanding these relationships crucial for investors and policymakers to predict stock price movements accurately. To this end, we examined the main stock market indices of G7 and MINT countries from 2012 to 2024, using a recent graph neural network (GNN) algorithm called multivariate time series forecasting with graph neural network (MTGNN). This method allows for considering complex spatio-temporal connections in multivariate time series. In the implementations, MTGNN revealed that the US and Canada are the most influential G7 countries regarding stock indices in the forecasting process, and Indonesia and T\"{u}rkiye are the most influential MINT countries. Additionally, our results showed that MTGNN outperformed traditional methods in forecasting the prices of stock market indices for MINT and G7 countries. Consequently, the study offers valuable insights into economic blocks' markets and presents a compelling empirical approach to analyzing global stock market dynamics using MTGNN.

\end{abstract}




\end{frontmatter}

\section{Introduction}
\label{sec:intro}
In the early 2010s, four countries attracted the attention of economists with their highly promising economies, and a new acronym, MINT: Mexico, Indonesia, Nigeria, and T\"{u}rkiye, emerged ~\citep{durotoye2014mint}. When the acronym was first suggested by Jim O'Neill, common characteristics of MINT countries were (i) geographical positions \textit{(Mexico sits next to the USA and belongs to the North American Free Trade Agreement (NAFTA), Indonesia lies at the centre of South East Asia, T\"{u}rkiye is connected to both the West and East, while Nigeria is on the coast of Africa surrounded by future trading partners)}, (ii) large populations \textit{(primarily under 30)}, (iii) rapid economic growth potential, (iv) a developing middle class, and (v) high levels of entrepreneurship ~\citep{Jim, Newland, Nagashybayeva}.  Within the last decade, the economic performance of MINT countries has been mixed, with some facing financial difficulties, including high inflation and political uncertainty that affect the investment climate ~\citep{Okoroafor}. In addition, according to the literature ~\cite{zhang2021stock,siddiqui2023assessing}, MINT countries as emerging economies may not only have been affected by domestic challenges in this decade but may also have been influenced by global conjuncture, particularly the developments in advanced economies, such as the group of seven countries.

The group of seven developed countries (G7), comprises the world's largest and
most advanced economies: Canada, France, Germany, Italy, Japan, the
United Kingdom, and the United States. They have a substantial impact on the development of global economic policies and trends.  As of 2023,  some of the economic properties of the G7 countries are (i) having the still largest share of the world's gross domestic product (GDP) \textit{(approximately 30\% in 2023, it was more than 43\% in the early 2000s)}, (ii) being an important player in global trade and investment due to multinational corporations, and (iii) having the most liquid and developed financial market centers that influence global capital flows and market dynamics ~\citep{World, United, Statista}. Due to these properties, G7 countries significantly impact emerging economies such as BRICS: Brazil, Russia, India, China, South Africa and MINT by shaping their economic policies, trade relations, and financial markets. Especially, the stock markets of emerging economies are often highly sensitive and closely tied to financial and economic developments in the G7 countries because financial markets are interconnected ~\citep{fratzscher2012capital, rey2015dilemma, acharya2020financial}. Therefore, understanding the links between developed and emerging markets is crucial for multiple stakeholders, including policymakers, investors, businesses, and researchers. To meet this crucial knowledge, in this study, one of the recent spatio-temporal graph neural networks (GNN) framework, known as MTGNN, was used to reveal these connections for the first time ~\citep{wu2020connecting}. The MINT and G7 countries' stock market indices were forecasted more precisely through these revealed interconnections by MTGNN.

Our study contributes to the current literature in a few important ways. First, the study introduces using of spatio-temporal GNN for economic blocks that considers them as a graph-network structure for the first time. Current economic approaches such as economic integration models do not explicitly conceptualize economic blocs as graph-based networks with evolving relationships over time. By using a spatio-temporal GNN framework, the study suggests a new methodological perspective for global economies. Second, from the point of empirical aspect, the study demonstrates how spatio-temporal GNN effectively capture complex and unknown interdependencies between economic blocks, in terms of their stock market indices. Leveraging the graph structure of stock indices and their interconnections, the study presents strong evidence as an improvement in prediction accuracy and robustness over some traditional, hybrid, and recent deep-learning approaches. For this reason, we metaphorically used the phrases of ``stock market telepathy'' and ``predicting secret conversations'' in the title to highlight GNN's ability to anticipate stock market movements and interdependencies with excellent accuracy.

As it is known, accurate stock price forecasting is a critical activity that supports decision-making across multiple domains, ensuring better financial outcomes and contributing to the stability and growth of the global economy. From this point of view, in addition to the contributing ways in the above, the study becomes significant when MINT countries' development could significantly impact international trade and economics in the years ahead, and the G7 countries' leadership will pursue, as well. In particular, understanding MINT countries' stock markets and seeing their predictability means both information about alternative investment destinations and the opportunity to create a portfolio with data on the functioning of the markets.

The rest of the paper is organized as follows: Section 2 reviews the existing literature on methods of predicting stock prices. Section 3 discusses the data and their properties. Section 4 provides a short review of spatio-temporal GNN methodology in particular that of MTGNN, and Section 5 presents empirical results. Finally, Section 6 addresses some conclusions with directions for future research.

\section{Literature review}
\label{sec:litreview}
The stock market plays a vital role in the economy of nations, serving as the primary platform for global capital exchange. Hence, the success of the stock market has a substantial impact on the overall state of the national economy. Investors in the stock market strive to maximize their earnings by analyzing market information and taking actions in response.  Investors might exploit the financial market using accurate models to forecast the stock price, influenced by macroeconomic elements and numerous other factors ~\citep{gao2020application}. 
Traditionally, the efficient market hypothesis also argues that future stock prices can be predicted using historical stock data ~\citep{fama1970efficient, shahi2020stock}. However, predicting stock index prices has long been a difficult task for professionals in the financial industry and related fields, mainly due to the presence of non-linearity, volatility, and noise characteristics. Thus, improving the precision of stock index price prediction and obtaining an accurate prediction is still a much-debated topic ~\citep{binkowski2018autoregressive}. In time, several traditional statistical models have been used to predict stock prices using historical data ~\citep{jarrett2011arima, tsai2012relationship, mensi2014global, sahoo2015stock, cakra2015stock, suharsono2017comparison, ma2018investor, izzeldin2019forecasting, tulcanaza2019determinants, ning2019stock}. 

In the last 20 years, the rise of computational intelligence has led to the development of advanced models \textit{(machine learning, deep-learning, and hybrid models)} for stock market forecasting. Instead of traditional statistical models that can only consider linear structures like autoregressive integrated moving average (ARIMA), vector autoregressive (VAR), plenty of new techniques such as Bayesian networks, fuzzy neural systems, genetic algorithms, recurrent neural networks (RNN), convolutional neural networks (CNN), and long-short term memory (LSTM) have been suggested by numerous researchers ~\citep{cheng2010hybrid, karazmodeh2013stock, chen2015hybrid, chong2017deep, hiransha2018nse, cao2020multiobjective, nikou2019stock, hargreaves2020stock, shahi2020stock, setiani2021prediction, pahlawan2021stock, alkhatib2022new, nasiri2023multi}. 

In addition to all these, in the last five years, the use of graph neural networks in stock price prediction has gained significant attention. ~\cite{deng2019knowledge} introduced a knowledge-driven temporal convolutional network (KDTCN) for stock trend prediction and explanation, emphasizing the importance of knowledge-driven events in predicting abrupt changes. ~\cite{long2020integrated} integrated deep learning and knowledge graph techniques to predict stock price trends in the Chinese stock exchange market. ~\cite{sawhney2020spatiotemporal} introduced the spatio-temporal hypergraph convolution network (STHGCN) for stock movement forecasting, highlighting its applications in quantitative trading and investment decision-making. Furthermore, ~\cite{sawhney2020deep} proposed a deep attentive learning architecture for stock movement prediction, leveraging financial data, social media, and inter-stock relationships through a graph neural network. ~\cite{wu2020connecting} discussed a novel approach to multivariate time series forecasting using graph neural networks (MTGNN). ~\cite{chen2021novel} introduced a graph convolutional feature based convolutional neural network (GC-CNN) model for stock trend prediction, demonstrating superior performance using Chinese stock data. ~\cite{hou2021st} developed the ST-trader model, a spatio-temporal deep neural network for modeling stock market movement, emphasizing the incorporation of inter-connections between firms to forecast stock prices. Lastly, ~\cite{sawhney2021exploring} explored the scale-free nature of stock markets and inter-stock correlations, proposing HyperStockGAT as a model for stock selection based on scale-free graph-based learning. These studies collectively highlight the advancements in utilizing graph neural networks for stock price prediction, emphasizing the importance of incorporating various data sources and methodologies to enhance prediction accuracy and decision-making in financial markets.

Following ~\cite{wu2020connecting}, this study for the first time employs a GNN framework-MTGNN, which can automatically extract the dependencies are not known in advance, to MINT and G7 countries' stock market indices. The success of the MTGNN method (i) at capturing complex, non-linear relationships over time between variables by representing them as nodes in a graph, (ii) obtaining accurate predictions ~\cite{wu2020connecting, cui2021metro, he2022multivariate, liu2022multivariate, jin2022multivariate, chen2022multi, chen2023multi}, and (iii) the thought of the countries in the MINT and G7 economic blocs naturally can be represented as a graph-network, (iv) the high possibility of these countries involving complex, non-linear economic interactions, and (v) these interactions between countries is critical to making accurate predictions caused to prefer the MTGNN over other methods.

\section{Data}
\label{sec:data}
We selected the main stock market indices as data to represent the stock market indices of MINT and G7 countries. As can seen from Table~\ref{tab:variable}, FTSE MIB index for Italy, BIST 100 index for T\"{u}rkiye, CAC 40 index for France, FTSE 100 index for UK, DAX PERFORMANCE index for Germany, S\&P 500 index for USA, S\&P/TSX index for Canada, IDX COMPOSITE index for Indonesia, IPC MEXICO index for Mexico, NIKKEI 225 index for Japan, and lastly NSE 30 index for Nigeria were chosen.  

The data we analyzed was the daily closing prices for the indices collected from
January 30, 2012 to August 14, 2024, accessed from Yahoo Finance ~\citep{yfinance} and Investing ~\citep{investing} websites. 
This date range was selected because the daily values of the Nigerian index NSE 30 is only avalaible from January 30, 2012. Further, the values of the BIST 100 index have been adjusted to reflect the change executed on July 27, 2020. This change removed two zeros from the values of the index on that date because it reached 100,000 points on June 13, 2017. Therefore, we divided the data preceding this date by 100 to adjust for this change. 

Table~\ref{tab:descriptive} depicts the descriptive statistics of the data. It
can be observed that all indices are non-normal, with the BIST 100 and NSE 30
(stock indices of members of MINT countries) indices being the most positively
skewed and leptokurtic.

\begin{figure*}
    \centering
    \includegraphics[width=1.0\textwidth]{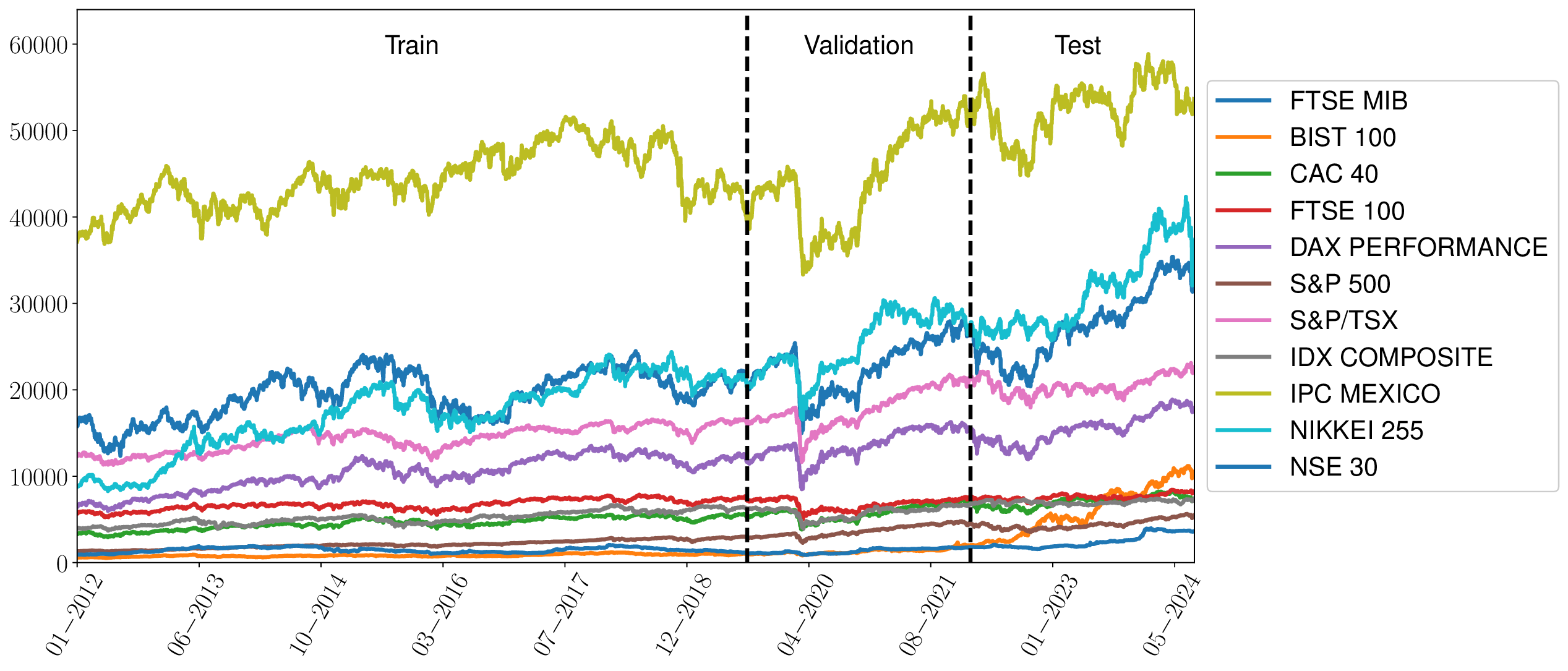}
    \caption{Daily closing price of the stock indices.}
    \label{fig:alldata}
\end{figure*}

All stock market indices values during that time were drawn in Figure~\ref{fig:alldata} for each country. According to Figure~\ref{fig:alldata}, although some dates are different, stock market movements seem generally similar. The decline in all of them, especially during the Covid period, is striking.

\begin{figure}[tbh]
\centering
  \includegraphics[width=0.475\textwidth]{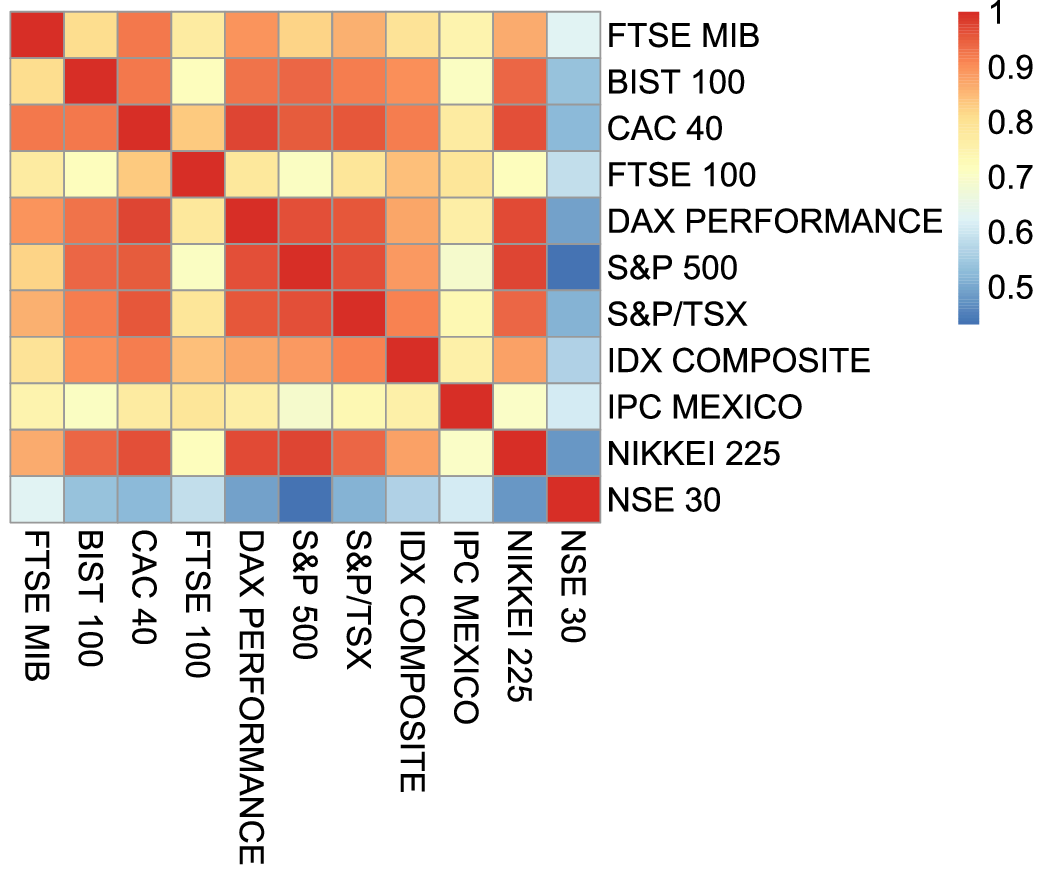}
  \caption{Correlation matrix produced by Spearman correlation analysis.} 
  \label{fig:cor2}
\end{figure}

\begin{figure}[tbh]
\centering
  \includegraphics[width=0.475\textwidth]{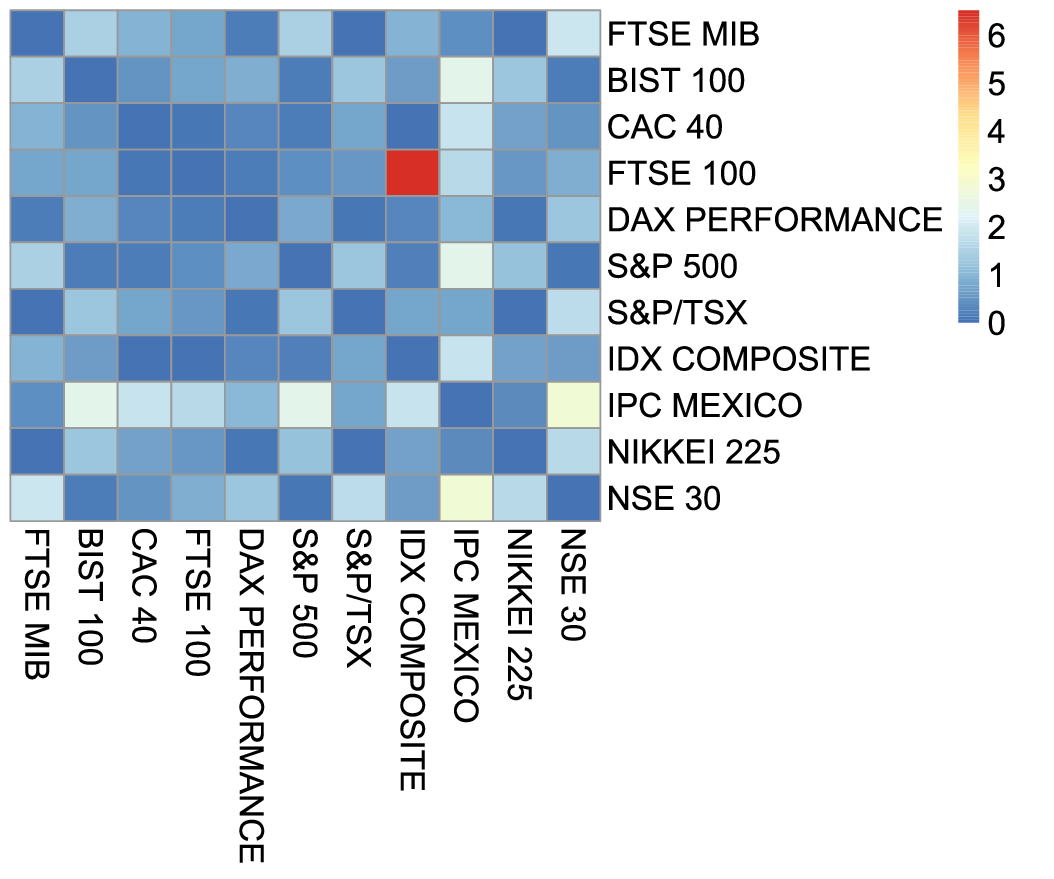}
  \caption{Distance matrix produced by dynamic time warping analysis.} 
  \label{fig:dist2}
\end{figure}

In addition to all data summaries, Spearman correlation analysis was also applied to the indices to examine their relationships. According to Figure~\ref{fig:cor2},  among MINT countries, Nigeria’s stock index was found to be least correlated with the other countries' indices, Mexico's index was second least correlated, and T\"{u}rkiye’s and Indonesia’s were most related. Spearman correlation analysis does not take into account the time-dependent 
relationships between the indices. A more appropriate method to analyze the
time-dependent relationships is dynamic time warping (DTW).  DTW is a robust
approach to determine a measure of distance which can be interpreted as a
measure of similarity between two time series, which may vary with time. The
primary concept of DTW is to calculate the distance by comparing corresponding
items in time series that are similar~\citep{dynamic}. Unlike traditional
distance measures, such as Euclidean distance, DTW can handle shifts and
distortions in the time axis and calculates a cumulative distance by considering
the minimum distance path through the cost matrix~\citep{muller2007dynamic}.
Lower DTW distances (close to $0$) indicate that the two time
series are more similar. Figure~\ref{fig:dist2} hence suggests that almost all 
indices show similarity to each other save for the interplay between the UK and
Indonesia.

\begin{table*}
    \caption{Variable explanations.}
      \label{tab:variable}
      \centering
    \scalebox{0.85}{
\begin{tabular}{@{}ll@{}}
\toprule
\textit{Variable} & \textit{Explanation} \\ \midrule
FTSE MIB   & Price performance of the 40 most-traded stock classes on the Borsa Italiana \\
BIST 100  & Price performance of the 100 largest companies on the Borsa Istanbul \\
CAC 40    & Price performance the 40 largest equities on the Euronext Paris\\
FTSE 100   &Price performance of the 100 most highly capitalised companies listed on the London Stock Exchange \\
DAX  PERFORMANCE   &Price performance of 30 biggest German companies that trade on the Frankfurt Exchange \\
S\&P 500    & Price performance of the largest 500 companies listed on stock exchanges in the United States \\
S\&P/TSX    & Stock market index representing roughly 70\% of the total market capitalization on the Toronto Stock Exchange \\
IDX COMPOSITE  & Index of all stocks listed on the Indonesia Stock Exchange \\
IPC MEXICO  &Weighted measurement index of most-traded 35 stocks on the Borsa Mexico\\
NIKKEI 225   &Price performance of the largest 225 companies on the Tokyo Stock Exchange\\
NSE 30   &Price performance of 30 largest companies on the Nigerian Stock Exchange \\ \bottomrule
\end{tabular}
}
\end{table*}

\begin{table*}
    \caption{Descriptive statistics for the main stock indices.}
    \label{tab:descriptive}
    \centering
    \scalebox{0.86}{
\begin{tabular}{@{}llllllllll@{}}
\toprule
\textit{Variable} & \textit{Source} & \textit{Size} & \textit{Mean} & \textit{Median} & \textit{Standard Deviation} & \textit{Minimum} & \textit{Maximum} & \textit{Skewness} & \textit{Kurtosis} \\ \midrule
FTSE MIB     & Yahoo Finance       & 4580       & 21732         & 21494           & 4422.896                    & 12358            & 35401            & 0.713             & 3.590             \\
BIST 100   & Yahoo Finance         & 4580       & 1969        & 977           & 2393.542                    & 541            & 11194          & 2.336             & 7.395             \\
CAC 40       & Yahoo Finance       & 4580       & 5308          & 5139            & 1216.689                    & 2929             & 8242             & 0.423             & 2.417             \\
FTSE 100     & Yahoo Finance       & 4580       & 6914          & 7004            & 644.245                     & 4994             & 8446             & -0.329            & 2.439             \\
DAX         & Yahoo Finance       & 4580       & 11990         & 12101           & 2865.310                     & 5976             & 18875            & 0.112             & 2.475             \\
S\&P 500     & Yahoo Finance       & 4580       & 2886          & 2680            & 1102.271                    & 1278             & 5644             & 0.526             & 2.149             \\
S\&P/TSX         & Yahoo Finance       & 4580       & 16280         & 15582           & 2957.038                    & 11310            & 23105            & 0.439             & 2.098             \\
IDX COMPOSITE        & Yahoo Finance       & 4580       & 5671          & 5779            & 961.285                     & 3697             & 7422             & -0.022            & 1.832             \\
IPC  MEXICO      & Yahoo Finance        & 4580       & 45960         & 45181           & 5118.520                     & 33338            & 58856            & 0.205             & 2.407             \\
NIKKEI 225    & Yahoo Finance      & 4580       & 21605         & 21103           & 7127.358                    & 8279             & 42344            & 0.376             & 2.909             \\
NSE 30      & Investing        & 4580       & 1635          & 1579            & 584.534                     & 872              & 3984             & 2.095             & 8.217             \\ \bottomrule
\end{tabular}
}
\end{table*}

\section{Methodology}
\label{sec:methodology}
In this section, we provided a brief background on GNN and MTGNN, which serves two purposes: (i) familiarize the reader with how they work to produce predictions and 
(ii) justify their use for price prediction of time series data.  We also briefly discuss the baseline methods we use for comparison. The range of baselines we consider comprises classical, hybrid, and deep-learning methods.

\subsection{Graph neural networks}
\label{ssec:gnn}

A graph is a tuple consisting of a set of nodes or vertices, where pairs of nodes are connected by 
edges or links. This data structure is used to describe the relationships between different
entities. Many real-world objects may naturally be described by constructing appropriate graph 
structures. For example, we can represent molecules by assigning different atoms to different nodes and different chemical bonds to different edges~\citep{prince2023understanding}. We can model transportation systems, such as road networks, railway systems, and flight routes, using graphs where nodes represent locations (cities, stations, airports), and edges represent the paths connecting them ~\citep{rahmani2023graph}.

GNNs are the preferred neural network topology when it is desired to generate representations of nodes
that actually depend on the structure of an underlying graph, as well as any feature information
the nodes might have~\citep{hamilton2020graph}. More recently, the literature has seen the emergence of a
special type of GNN, called spatio-temporal graph neural networks, that are designed to deal with
multivariate time series. They have first been applied to the task of traffic
prediction~\citep{chen2020multi,li2017diffusion,wu2019graph,yu2017spatio,zheng2020gman}. While the vanilla 
GNN architecture is responsible to resolve spatial dependencies among nodes, the temporal dependencies 
are resolved by the use of recurrent neural networks~\citep{li2017diffusion,seo2018structured} or 1D 
convolutions~\citep{yan2018spatial,yu2017spatio}. Spatio-temporal GNNs take multivariate time series along
with an underlying graph structure that describes the relationship among variables as inputs. Unfortunately, 
raw time-series data is typically not presented with a graph structure that describes the dependence of
variables on each other. This structure also needs to be learned from raw time series data.

\subsubsection{Formulating Multivariate Time Series with GNN}
Following the development in~\cite{wu2020connecting}, we let $\bm{z}_t \in \mathbb{R}^N$ to denote the 
values of an $N$-dimensional multivariate time series at the time index $t$. Daily stock index observations
for $N$ countries are arranged in a sequence of $P$ time steps $\mc{X} \supseteq \bm{X} = 
\left\{\bm{z}_{t_1}, \bm{z}_{t_2}, \ldots \bm{z}_{t_P}\right\}$. The goal is to predict a sequence of
future values $\mc{Y} \supseteq \bm{Y} = \left\{\bm{z}_{t_{P+1}}, \bm{z}_{t_{P+2}}, \ldots \bm{z}_{t_{P+Q}}\right\}$. 
This goal is to be achieved by constructing a map $f: \mc{X} \rightarrow \mc{Y}$ as a spatio-temporal 
graph neural network by minimizing absolute loss with regularization $\ell_2$.

As a reminder, formal definitions of most important graph theory concepts are presented below.

\begin{defn}[Graph]
   A graph $\mc{G} = (\mc{V}, \mc{E})$ is a data structure consisting of a set of \textit{nodes} $\mc{V}$ 
   and \textit{edges} $\mc{E}$. The total number of nodes in a graph $\mc{G}$ is denoted by $N$.
\end{defn}

\begin{defn}[Neighborhood]
   Suppose $u, v \in \mc{V}$ and there exists an edge $e = (v, u) \in \mc{E}$ pointing from $u$ to $v$. 
   We say that the set of all such $u \in \mc{V}$ constitutes the neighborhood of the node $v$, denoted by $\mc{N}(v) = \{ u \in \mc{V}: (v, u) \in \mc{E}\}$.
\end{defn}

\begin{defn}[Adjacency Matrix]
    The adjacency matrix $\bm{A} \in \mathbb{R}^{N \times N}$ is constructed such that $\bm{A}_{ij} > 0$ if 
    and only if there is an edge pointing from $v_i$ to $v_j$. All other entries are set to zero.
\end{defn}

In MTGNN, each variable in a multivariate time series data is represented as a node in a graph. The edges in the graph represent how the information flows between the nodes.

\subsubsection{MTGNN Model Architecture}

We borrow the MTGNN model from~\cite{wu2020connecting}. In this subsection, we briefly touch upon the 
most important aspects of this model and invite the reader to the original paper for a more detailed 
exposition.

\begin{figure}[tbh]
  \centering
  \includegraphics[width=0.48\textwidth]{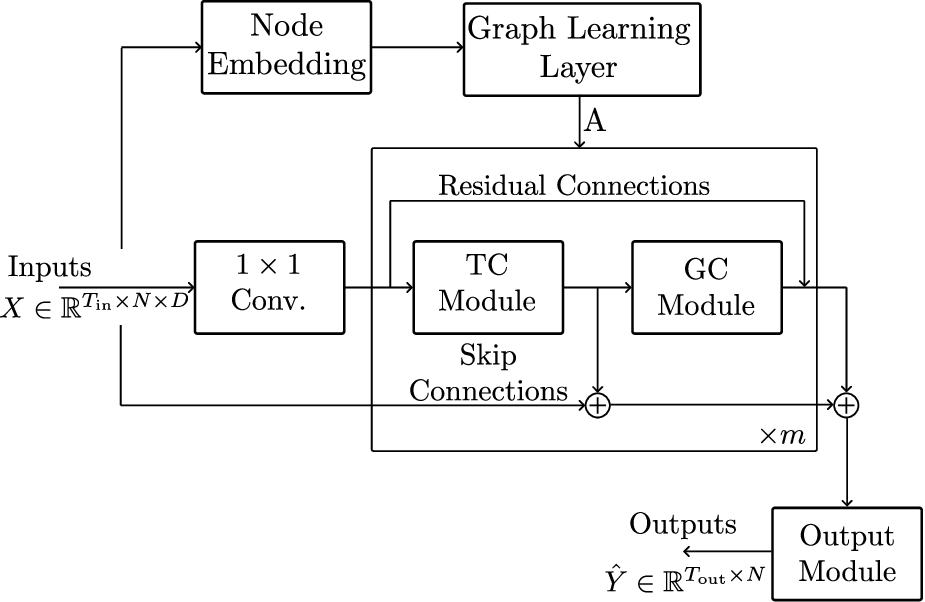}
  \caption{The architecture of the spatio-temporal graph neural network-MTGNN.} 
  \label{fig:gnn_architecture}
\end{figure}

The spatio-temporal GNN model operates by embedding the node features in a higher-dimensional space
\textit{($\textit{40}$ is used in this study)}. These embeddings are then fed into a graph learning layer, which automatically
trains the learnable parameters to compute the graph structure in conjunction with the remaining learnable
parameters for prediction. The graph learning layer outputs the current belief on the adjacency matrix at 
each learning step. This adjacency matrix and the node features are subsequently fed into a sequence of 
interleaved $m$ graph convolution and $m$ temporal convolution modules. Residual connections are added from
the inputs of the temporal convolution to the outputs of the graph convolution modules. Furthermore, 
each temporal convolution module is followed by skip connections. The residual and skip connections are 
inserted in order to fight the problem of gradient vanishing. Finally, the final outputs are computed 
by the output modules, which projects the hidden features to the desired output dimension. 
Figure~\ref{fig:gnn_architecture} provides a demonstration of the full set-up. Readers who are interested 
in more details on the neural network architecture are invited to peruse~\citep{wu2020connecting}.

\subsection{Baseline Forecasting Methods for Comparison}
\label{ssec:multi_ts}
We establish baselines to compare and contrast MTGNN by measuring the performance of four other 
prominent methods used to process multivariate time series data. One is the classical autoregressive (AR) model, 
while the rest are neural-network-based methods: autoregressive-multilayer perception (VAR-MLP), 
recurrent neural network-gated recurrent unit (RNN-GRU) and temporal convolutional network (TCN).

An autoregressive model assumes that the current value of a time series is a function of its past values. In other words, AR model is when a value from a time series is regressed on previous values from that same time series. In this regression model, the response variable in the previous time period has become the predictor and the errors have usual assumptions about errors in a simple linear regression model. The order of an autoregression is the number of immediately preceding values in the series that are used to predict the value at the present time ~\citep{penn}. 

The vector autoregressive (VAR) model is a well-known statistical model used to capture linear interdependencies among multiple time series. It has been widely applied in econometrics for modeling and forecasting systems where multiple variables influence each other over time ~\citep{lutkepohl2005new}. However, VAR models assume linear relationships, which can be limiting in cases where the underlying data exhibits complex, nonlinear dynamics. To address this limitation, the VAR-MLP model incorporates the MLP, a type of artificial neural network known for its ability to learn and approximate nonlinear functions. By combining VAR with MLP, the VAR-MLP model leverages the strengths of both methods: it captures the linear interdependencies among multiple time series using VAR, while the MLP component models the potential nonlinear relationships that VAR cannot. This hybrid approach allows for more accurate modeling and forecasting in systems where both linear and nonlinear interactions are significant, as demonstrated in empirical studies across various domains, including finance and economics ~\citep{zhang2003time, zivot2006modeling}. The flexibility of VAR-MLP makes it particularly effective in complex systems where traditional linear models may fall short, thus providing a comprehensive framework for time series analysis. 

RNN is a class of neural networks designed to handle sequential data by maintaining a hidden state that captures information from previous inputs ~\citep{hochreiter1997long}. However, RNN suffers from issues like vanishing gradients, which limit their ability to learn long-term dependencies. To address this, the gated recurrent unit (GRU) was introduced as an improved variant of the RNN ~\citep{cho2014learning}. The GRU uses gating mechanisms—namely, an update gate and a reset gate—to control the flow of information, making it more efficient at capturing dependencies over longer sequences while reducing the complexity ~\citep{chung2014empirical}. This innovation allows GRU to perform better than traditional RNNs in various sequential modeling tasks by mitigating the vanishing gradient problem and simplifying the model structure.

TCN is a type of neural network architecture designed for sequence modeling tasks, offering an alternative to RNN and LSTM networks. TCN leverages the strengths of CNN by using 1D convolutions to process sequential data, enabling them to capture temporal dependencies over long sequences efficiently. Unlike RNN, TCN does not suffer from issues like vanishing gradients and can handle much longer input sequences due to their ability to employ dilated convolutions, which exponentially increase the receptive field of the network without a corresponding increase in computational cost ~\citep{bai2018empirical}. Additionally, TCN uses causal convolutions, ensuring that the output at any time step is only influenced by past inputs, maintaining the temporal order of the data ~\citep{oord2016wavenet}. This structure allows TCN to outperform traditional RNN and LSTM in various tasks, such as time series forecasting, due to their stable training dynamics and ability to capture long-range dependencies ~\citep{bai2018empirical}.

\subsection{Evaluation metrics}
\label{ssec:metric}

The metrics we used to measure the performance of applied algorithms in the study are relative squared error (RSE), root mean squared error (RMSE), mean absolute error (MAE), and mean absolute percentage error (MAPE). The mathematical expressions for these metrics are given below.
\begin{align}
    \operatorname{RSE} &= \frac{\sum_{i=1}^N \left(y_i - \hat{y}_i\right)^2}{\sum_{i=1}^N \left(y_i - \bar{y}\right)^2}, \\
    \operatorname{RMSE} &= \sqrt{\frac{\sum_{i=1}^N \left(y_i - \hat{y}_i\right)^2}{N}}, \\
    \operatorname{MAE} &= \frac{\sum_{i=1}^N \abs{y_i - \hat{y}_i}}{N}, \\
    \operatorname{MAPE} &= \frac{1}{N}\sum_{i=1}^N \abs{\frac{y_i - \hat{y}_i}{y_i}},
\end{align}
where $\hat{y}_i$ is the value predicted by the algorithm for observation $i$ (out of $N$ observations), $y_i$ is 
the actual value, and $\bar{y}$ is the average of all target values. For a perfect fit, each of these 
measures would assume the value $0$. Hence, they range from $0$ to $\infty$ with $0$ corresponding to 
the ideal.

\section{Results and Discussion}
\label{sec:results}
This section presents the findings derived from applying MTGNN, and other baseline forecasting methods \textit{(AR, VAR-MLP, RNN-GRU, TCN)} in forecasting the stock indices of MINT and G7 countries. Before employing all analyses, values of data were transformed to natural logarithm, and the initial $60\%$ of 
the dataset used for training the models, the subsequent $20\%$ of the data was used for validation, and the last $20\%$ of the data, starting from January 2022 and ending on August 14, 2024 were used for testing purposes. 

All the hyperparameters of the MTGNN used to obtain the results are summarized in Table~\ref{tab:hyperparams}. 

\begin{table}[bt]
    \caption{The hyperparameters of the MTGNN used to train and infer the forecasts.}
    \label{tab:hyperparams}
    \centering
    \begin{tabular}{ *3l }           \toprule
    \emph{Hyperparameter}  & \emph{Value} \\ \cmidrule(lr){1-2}
    Convolution depth & $2$ \\
    Loss & $\ell_1$ \\
    Dropout & $0.3$ \\
    Convolutional channels & $16$ \\
    Residual channels & $16$ \\
    Skip channels & $32$ \\
    Batch size & $8$ \\ 
    Epochs & $30$ \\
    \bottomrule
    \hline
    \end{tabular}
\end{table}

\begin{figure}[tbh]
  \centering
  \includegraphics[width=0.48\textwidth]{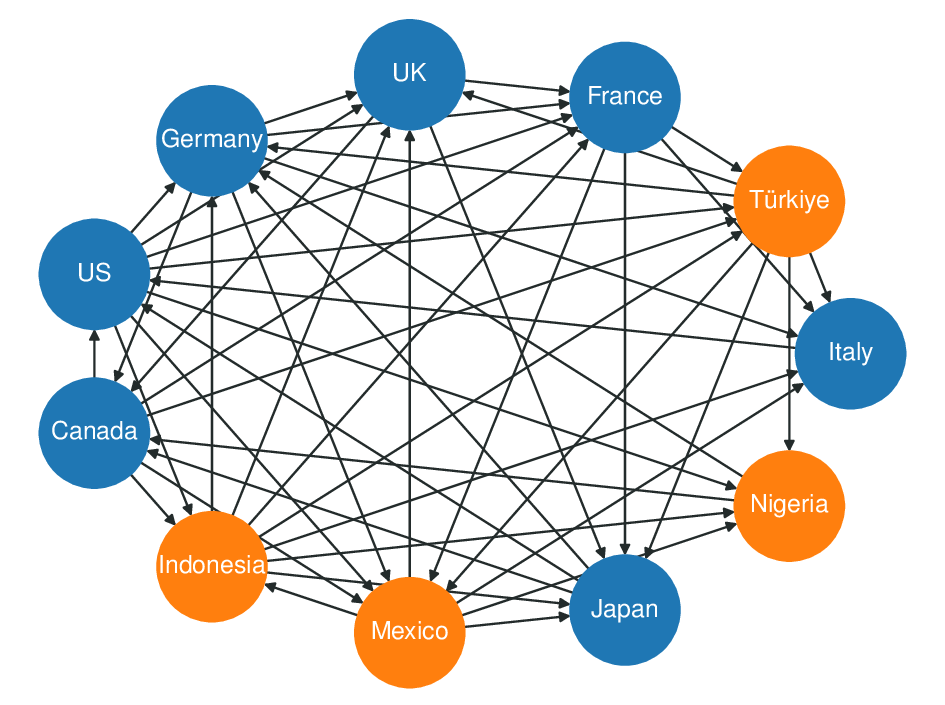}
  \caption{The connectivity of the countries' stock indices deduced by the MTGNN.} 
  \label{fig:graph}
\end{figure}

After training on the first $60\%$ of the data as shown in
Figure~\ref{fig:alldata}, MTGNN produces the underlying graph structure that minimizes the $\ell_1$ loss as in Figure~\ref{fig:graph}.  In the graph, each node represents the stock indices of a country, and each edge represents a link between a pair of stock indices. As shown in Table~\ref{tab:hyperparams}, the depth of the graph convolutions is selected to be $2$. This means that each node had their feature vector updated by considering the current information residing in its $2$-hop neighborhood \textit{(i.e., its neighbors and the neighbors of its neighbors)}. 
The constructed graph in Figure~\ref{fig:graph} can be interpreted up to 2-hop neighbors have a direct effect on the outcome of each node, while the remaining nodes (variables) only have an indirect effect in predicting the future values of a particular node. Notice that this graph is constructed by performing and end-to-end stochastic gradient descent (more precisely, the Adam optimizer), on the losses accrued
between the predicted and the actual values. This means that by no means do we claim that the constructed graph in Figure~\ref{fig:graph}, is a causal one. It is the graph that the optimization has constructed such that the resulting loss
is minimized. 

The adjacency matrix, corresponding to the graph in Figure~\ref{fig:graph} is given by \textbf{A} in Equation~\eqref{eq:adjacency}. The matrix provides a visual representation of the connections \textit{(edges)} and their strengths between the countries' stock indices. Zeros in the matrix demonstrate no connection between pairs of stock indices. Performing sums over the columns of the adjacency
matrix~\eqref{eq:adjacency} reveals that if only $1$-hop neighborhoods are considered, the US, Germany, and Canada are the most influential stock indices among G7 countries with $7$, $5$, and $5$ out-degree connections, respectively, while among MINT countries, Indonesia and T\"{u}rkiye are the most influential stock indices with $7$ and $6$ out-degree connections, respectively, in the forecasting process. Similarly, the
out-degrees of the $2$-hop neighborhoods can be computed by considering the column sums of the matrix $\bm{A} + \bm{A}^2$. In this case, the US and Canada
are the most influential stock indices among G7 countries with $39$ and $34$ out-degree connections, respectively, while among MINT countries, Indonesia and T\"{u}rkiye are the most influential stock indices with $31$ and $25$ out-degree connections, respectively, in the forecasting process.

The results in Table~\ref{tab:comparison} show that the performance of the MTGNN, and other baseline comparison methods, AR, VAR-MLP, RNN-GRU, and TCN, in terms of RSE, RMSE, MAE, and MAPE evaluation metrics. The best results are bolded, and the second-best results are underlined in Table~\ref{tab:comparison}. It can be observed that MTGNN is excellent across all the stock indices. The table
shows that the second-best results vary across the board. While VAR-MLP achieves this honor for the indices of five countries \textit{(France, US, Canada, Indonesia, and Mexico)}, TCN is the second-best model for three countries \textit{(T\"{u}rkiye, Japan, and Nigeria)}, and RNN-GRU is the second-best model for two countries \textit{(Italy and Germany)}.

\setcounter{MaxMatrixCols}{11}
\small
\begin{equation}
\label{eq:adjacency}
\bm{A} = \begin{bmatrix}
0 & 1 & 1 & 0 & 1 & 0 & 0 & 0.98 & 1 & 0 & 0 \\ 
0 & 0 & 1 & 0 & 0 & 1 & 1 & 1 & 0 & 0 & 0 \\
0 & 0 & 0 & 1 & 0.96 & 1 & 1 & 1 & 0 & 0 & 0 \\
0 & 1 & 0 & 0 & 1 & 1 & 0 & 1 & 1 & 0 & 0 \\
0 & 1 & 0 & 0 & 0 & 1 & 0 & 1 & 0 & 1 & 0.99 \\
1 & 0 & 0 & 0 & 0 & 0 & 1 & 0 & 0 & 1 & 0 \\
0 & 0 & 0 & 1 & 0.9 & 0 & 0 & 0 & 0 & 1 & 1 \\ 
0 & 0 & 0 & 0 & 0 & 1 & 1 & 0 & 1 & 0 & 0 \\
0 & 1 & 1 & 0 & 1 & 1 & 1 & 0 & 0 & 0 & 0 \\
0 & 0.81 & 1 & 1 & 0 & 0 & 0 & 1 & 1 & 0 & 0 \\
0 & 1 & 0 & 0 & 0 & 1 & 0 & 1 & 1 & 0 & 0
\end{bmatrix}
\end{equation}
\normalsize

\begin{figure*}[tbh]
  \centering
  \includegraphics[width=0.97\textwidth]{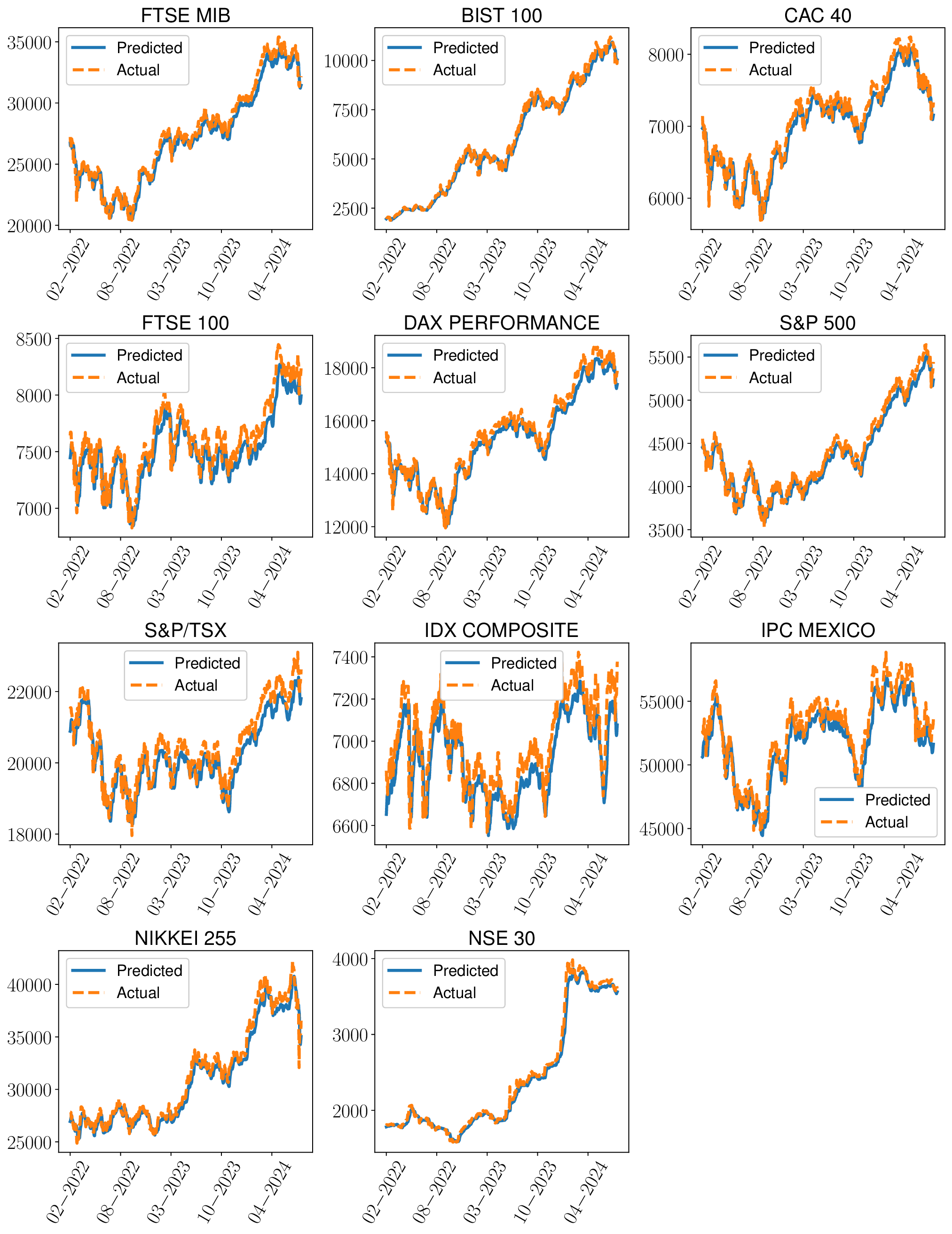}
  \caption{The predictions obtained MTGNN predictor.} 
  \label{fig:predictions}
\end{figure*}

For completeness, we have provided plots of the MTGNN's predictions of all closing prices of stock indices of MINT and G7 countries for the test dataset in Figure~\ref{fig:predictions}. The performance of the MTGNN is not only satisfactory in terms of its RSE, RMSE, MAE, and MAPE performances, but is also visually pleasing, as can be observed by perusing each plot in the figure. 

\begin{rem}
The MTGNN algorithm normalizes the data by subtracting from each column its individual mean and dividing by its individual standard deviation. All data transformations \textit{(natural logarithm and normalization)} implemented are inverted for the construction of the plots in Figure~\ref{fig:predictions}.
\end{rem}

\setlength{\tabcolsep}{18pt}
\begin{table*}[bt]
    \caption{Comparison of performances of various algorithms from the literature.}
    \label{tab:comparison}
    \centering
    \scalebox{0.89}{
    \begin{tabular}{ *6l }           \toprule
    \textit{Index}& \textit{Algorithm} & \textit{RSE} & \textit{RMSE} & \textit{MAE} & \textit{MAPE} \\  
    \cmidrule(lr){1-6}
    \multirow{5}{*}{\hspace{-5mm}\minitab[l]{FTSE MIB \\ (Italy)}} & AR & $3.503$ & $0.276$ & $0.242$ & $2.4\%$  \\
    & VAR-MLP & $1.066$ & $0.152$ & $0.125$ & $1.2\%$ \\
    & RNN-GRU & $\underline{1.005}$ & $\underline{0.148}$ & $\underline{1.005}$ & $\underline{1.1}\%$\\
    & TCN & $1.545$ & $0.183$ & $0.174$ & $1.7\%$\\
    & MTGNN & $\textbf{0.026}$ & $\textbf{0.024}$ & \textbf{0.020} & $\textbf{0.2}\%$\\
    \midrule
    \multirow{5}{*}{\hspace{-5mm}\minitab[l]{BIST 100 \\ (T\"{u}rkiye)}} & AR & $10.294$ & $1.672$ & $1.600$ & $18.4\%$ \\
    & VAR-MLP & $1.487$ & $0.633$ & $0.558$ & $6.4\%$ \\
    & RNN-GRU & $0.890$ & $0.492$ & $0.428$ & $4.9\%$\\
    & TCN & $\underline{0.439}$ & $\underline{0.346}$ & $\underline{0.331}$ & $\underline{3.8}\%$ \\
    & MTGNN & $\textbf{0.081}$& $\textbf{0.042}$ & $\textbf{0.033}$ & $\textbf{0.4}\%$ \\
    \midrule
    \multirow{5}{*}{\hspace{-5mm}\minitab[l]{CAC 40 \\ (France)}} & AR & $6.939$ & $0.237$ & $0.227$ & $2.6\%$ \\
    & VAR-MLP & $\underline{0.980}$ & $\underline{0.089}$ & $\underline{0.076}$ & $\underline{0.9}\%$ \\
    & RNN-GRU & $5.589$ & $0.213$ & $0.184$ & $2.1\%$\\
    & TCN & $2.246$ & $0.135$ & $0.133$ & $1.5\%$\\
    & MTGNN & $\textbf{0.045}$ & $\textbf{0.020}$ & $\textbf{0.017}$ & $\textbf{0.2}\%$\\
    \midrule
    \multirow{5}{*}{\hspace{-5mm}\minitab[l]{FTSE 100 \\ (UK)}} & AR & $\underline{0.872}$ & $\underline{0.040}$ & $\underline{0.031}$ & $\underline{0.3}\%$\\
    & VAR-MLP & $1.562$ & $0.053$ & $0.046$ & $0.5\%$ \\
    & RNN-GRU & $6.76$ & $0.111$ & $0.093$ & $1.0\%$\\
    & TCN & $10.208$ & $0.136$ & $0.133$ & $1.5\%$\\
    & MTGNN & $\textbf{0.044}$ & $\textbf{0.018}$ & $\textbf{0.016}$ & $\textbf{0.2}\%$\\
    \midrule
    \multirow{5}{*}{\hspace{-5mm}\minitab[l]{DAX PERFORMANCE \\ (Germany)}} & AR & $2.891$ & $0.192$ & $0.175$ & $1.8\%$ \\
    & VAR-MLP & $1.017$ & $0.114$ & $0.092$ & $0.9\%$ \\
    & RNN-GRU & $\underline{0.405}$ & $\underline{0.072}$ & $\underline{0.047}$ & $\underline{0.5}\%$\\ 
    & TCN & $2.706$ & $0.169$ & $0.124$ & $1.9\%$\\
    & MTGNN & $\textbf{0.032}$ & $\textbf{0.023}$ & $\textbf{0.020}$ & $\textbf{0.2}\%$\\
    \midrule
    \multirow{5}{*}{\hspace{-5mm}\minitab[l]{S\&P 500 \\ (US)}} & AR & $10.008$ & $0.357$ & $0.346$ & $4.1\%$ \\
    & VAR-MLP & $\underline{0.965}$ & $\underline{0.111}$ & $\underline{0.092}$ & $\underline{1.1}\%$ \\
    & RNN-GRU & $2.879$ & $0.192$ & $0.157$ & $1.9\%$\\
    & TCN & $2.241$ & $0.169$ & $0.163$ & $1.9\%$ \\
    & MTGNN & $\textbf{0.031}$ & $\textbf{0.021}$ & $\textbf{0.018}$ & $\textbf{0.2}\%$\\
    \midrule
    \multirow{5}{*}{\hspace{-5mm}\minitab[l]{S\&P/TSX \\ (Canada)}}& AR & $16.568$ & $0.209$ & $0.203$ & $2.0\%$ \\
    & VAR-MLP & $\underline{1.104}$ & $\underline{0.054}$ & $\underline{0.043}$ & $\underline{0.4}\%$ \\
    & RNN-GRU & $51.66$ & $0.368$ & $0.337$ & $3.4\%$\\
    & TCN & $5.929$ & $0.125$ & $0.124$ & $1.3\%$\\
    & MTGNN & $\textbf{0.024}$ & $\textbf{0.020}$ & $\textbf{0.017}$ & $\textbf{0.2}\%$\\
    \midrule
    \multirow{5}{*}{\hspace{-5mm}\minitab[l]{IDX COMPOSITE\\ (Indonesia)}} & AR & $2.812$ & $0.050$ & $0.040$ & $0.5\%$ \\
    & VAR-MLP & $\underline{1.111}$ & $\underline{0.031}$ & $\underline{0.027}$ & $\underline{0.3}\%$ \\
    & RNN-GRU & $17.536$ & $0.124$ & $0.117$ & $1.3\%$\\
    & TCN & $66.671$ & $0.241$ & $0.231$ & $2.61\%$ \\
    & MTGNN & $\textbf{0.037}$ & $\textbf{0.016}$ & $\textbf{0.014}$ & $\textbf{0.2}\%$\\
    \midrule
    \multirow{5}{*}{\hspace{-5mm}\minitab[l]{IPC MEXICO\\ (Mexico)}} & AR & $6.134$ & $0.152$ & $0.140$ & $1.3\%$ \\
    & VAR-MLP & $\underline{0.987}$ & $\underline{0.061}$ & $\underline{0.048}$ & $\underline{0.4}\%$ \\
    & RNN-GRU & $1.526$ & $0.076$ & $0.068$ & $0.6\%$\\
    & TCN & $7.981$ & $0.173$ & $0.163$ & $1.5\%$\\
    & MTGNN & $\textbf{0.021}$ & $\textbf{0.022}$ & $\textbf{0.018}$ & $\textbf{0.2}\%$\\
    \midrule
    \multirow{5}{*}{\hspace{-5mm}\minitab[l]{NIKKEI 225 \\ (Japan)}} & AR & $5.273$ & $0.329$ & $0.315$ & $3.0\%$ \\
    & VAR-MLP & $1.597$ & $0.181$ & $0.160$ & $1.6\%$ \\
    & RNN-GRU & $1.654$ & $0.184$ & $0.148$ & $1.4\%$\\
    & TCN & $\underline{0.298}$ & $\underline{0.078}$ & $\underline{0.075}$ & $\underline{0.7}\%$\\
    & MTGNN & $\textbf{0.025}$ & $\textbf{0.025}$ & $\textbf{0.020}$ & $\textbf{0.2}\%$\\
    \midrule
    \multirow{5}{*}{\hspace{-5mm}\minitab[l]{NSE 30 \\ (Nigeria)}} & AR & $6.526$ & $0.747$ & $0.700$ & $8.9\%$ \\
    & VAR-MLP & $1.190$ & $0.319$ & $0.292$ & $3.8\%$ \\
    & RNN-GRU & $0.491$ & $0.205$ & $0.128$ & $1.6\%$\\
    & TCN & $\underline{0.045}$ & $\underline{0.062}$ & $\underline{0.057}$ & $\underline{0.7}\%$\\
    & MTGNN & $\textbf{0.013}$ & $\textbf{0.026}$ & $\textbf{0.018}$ & $\textbf{0.2}\%$ \\
    \bottomrule
    \hline
    \end{tabular}
    }
    \begin{tablenotes}
      \small
      \centering
      \item The best results are bolded, and the second best results are underlined.
    \end{tablenotes}
\end{table*}

\section{Conclusions}
\label{sec:conclusion}

While the stock markets of developed countries are still dominant in international markets, the influence of emerging economies has been growing recently. This growth has led economists to propose new economic blocs, such as MINT countries. Although MINT countries are in a growth phase, they are quite open to being influenced by advanced economies like G7. In particular, emerging economies' stock markets are frequently highly sensitive and closely linked to financial and economic developments in the G7 countries due to the interlinked nature of financial markets. Consequently, generating forecasts for stock price movement is crucial for (i) investors to manage risks and optimize their portfolios, (ii) stakeholders to guide policies that promote global financial stability, and (iii) policymakers to strategically design fiscal policies that stimulate stock market growth while understanding the connections between developed and emerging markets. In this vein, the stock market indices of G7 countries and MINT countries, representing developed and emerging economies, respectively, were discussed and predicted in the study between 2012 and 2024 with the help of graph neural networks. From this aspect, the study contributes significantly to understanding the financial interactions between the G7 and the MINT countries providing new empirical evidence.

In the study, a graph neural network architecture called MTGNN was applied for the first time to predict the closing prices of the main stock market indices of G7 and MINT economic blocs. This methodological approach allows for considering not only temporal connections but also spatial connections in multivariate time series. By accounting for the complex interconnections between countries, which are not known in advance, the forecasting process is enhanced, leading to improved prediction accuracy. As a result of the implementations, MTGNN revealed that the US and Canada are the most influential G7 countries regarding stock indices in the forecasting process, and Indonesia and T\"{u}rkiye are the most influential MINT countries in these economic blocks. Particularly, following the behaviors of aforesaid countries' stock markets, provides insights into other MINT and G7 countries' stock markets. In addition to these, the MTGNN model achieved excellent accuracy with minimum error amounts in stock price predictions compared to other baseline methods such as AR, VAR-MLP, RNN-GRU, and TCN. This improved predictive capability is particularly beneficial for emerging markets, which are often less stable and more sensitive to domestic and global economic conditions and, therefore, difficult to predict. 

No doubt, generating forecasts for stock price movements is advantageous for investors, investment managers, and policy-makers engaged in stock market prediction. New artificial intelligence-based deep-learning methods, such as GNN and its derivates, provided forecasts with unprecedented accuracy even in emerging economies like MINT, which have less stable economic environments and are more sensitive to domestic and global situations than developed countries. This study not only demonstrates the effectiveness of MTGNN in stock market forecasting but also lays the foundation for broader applications of GNN in financial analytics, offering valuable insights into global market dynamics. Future studies should be encouraged to employ these methods more frequently for better predictions in the economics and finance fields.

\section*{Disclosure Statement}
No potential conflict of interest was reported by the author.

\section*{Funding}
\label{sec:funding}
The author did not receive support from any organization for this study.


\clearpage

\bibliography{sample}

\end{document}